\documentclass[twocolumn,prl]{revtex4-1}
\usepackage{epsfig}

\newcommand{\ben}{\begin{enumerate}}
\newcommand{\een}{\end{enumerate}}

\newcommand{\<}{\langle}
\renewcommand{\>}{\rangle}
\newcommand{\lp}{\left(}
\newcommand{\rp}{\right)}
\newcommand{\lb}{\left[}
\newcommand{\rb}{\right]}

\newcommand{\beq}{\begin{equation}}
\newcommand{\eeq}{\end{equation}}
\newcommand{\bea}{\begin{eqnarray}}
\newcommand{\eea}{\end{eqnarray}}

\usepackage{bm}
\usepackage{amsmath}
\usepackage{amsthm}
\usepackage{amsfonts}

\begin{document}
\title{Dynamical Maximum Entropy Approach to Flocking}
\author{Andrea Cavagna$^{1,2,3}$, Irene Giardina$^{1,2,3}$, Francesco Ginelli$^{4}$, Thierry
  Mora$^{5}$, Duccio Piovani$^{2}$, Raffaele Tavarone$^{2}$ and Aleksandra M. Walczak$^{6}$}
\affiliation{$^1$ Istituto Sistemi Complessi, Consiglio Nazionale
  delle Ricerche, UOS Sapienza, Rome, Italy}
\affiliation{$^2$ Dipartimento di Fisica, Universit\`a\ Sapienza,
  Rome, Italy}
\affiliation{$^3$ Initiative for the Theoretical Sciences, The
  Graduate Center, The City University of New York, New York}
\affiliation{$^4$ SUPA, Institute for Complex Systems and Mathematical Biology, King's
College, University of Aberdeen, Aberdeen, UK}
\affiliation{$^5$ Laboratoire de physique statistique, CNRS,
  UPMC and
  \'Ecole normale sup\'erieure, Paris, France}
\affiliation{$^6$ Laboratoire de physique th\'eorique, CNRS,
  UPMC and \'Ecole normale sup\'erieure, Paris, France}
\date{\today}
\linespread{1}

\begin{abstract}

We derive a new method to infer
from data the out-of-equilibrium alignment dynamics of collectively
moving animal groups, by considering the maximum entropy distribution consistent with temporal
and spatial correlations of flight direction. When bird neighborhoods evolve rapidly, this
dynamical inference correctly learns the parameters of the model,
while a static one relying only on the spatial correlations
fails. When neighbors change slowly and detailed balance
is satisfied, we recover the static procedure. We demonstrate the
validity of the method on simulated data. The approach is
applicable to other systems of active matter.

\end{abstract}

\maketitle

Flocking, the highly coordinated motion displayed by large groups of birds, 
has attracted much attention over the last twenty years as a prototypical example of out-of-equilibrium collective behavior.
It has been suggested that flocking is an emergent phenomenon resulting from mutual alignment
of velocities between neighboring birds, much like the spontaneous symmetry breaking towards a magnetized state exhibited
by ferromagnetic spins at low temperatures. Although this idea has been extensively
studied from a theoretical view point
\cite{block1, TonerTu, Ginelli:2010p12925}, only recently have advances in the 3D imaging of large flocks of starlings
\cite{Cavagnafriends} 
given empirical grounds supporting this picture.
Interactions between individuals in the flock were
shown to be topological and local
\cite{Ballerini:2008p679}, leading to the global ordering of flight
orientations and  scale-free correlation functions
\cite{Cavagna:2010p7977}.
The analogy with ferromagnetic
systems was made explicit by the quantitative inference of spin
models from empirical data using the principle of maximum entropy \cite{Bialek:2012p12539,Bialek:2013p13047}.
These analyses have focused on the steady state behaviour of flocks,
by examining the flock configurations as 
drawn from a given statistical ensemble.
This approach allows for an effective equilibrium-like
description, without having to make detailed assumptions about the microscopic rules governing flock
behaviour. Yet it is an incomplete picture as it does not take into account the dynamical,
out-of-equilibrum nature of the process.

The major difference between flocks and  equilibrium spin systems is
that birds are like active particles, constantly moving 
within the flock along the direction given by their  ``spin'',
exchanging local interaction partners, thus extending their effective
interaction range, and also breaking detailed balance.
This qualitative difference between equilibrium spins and out-of-equilibrium active particles can dramatically
affect the thermodynamic properties of the system, including the
existence of an ordered phase in two dimensions, and the value
of the critical exponents \cite{TonerTu}. One can thus naively 
interpret the parameters of static descriptions of flocks
as a renormalized version of some underlying and unknown out-of-equilibrium dynamical model. 

In this paper we propose a general framework for learning the
features of the out-of-equilibrium dynamics directly from data, while making minimal
assumptions about the specific microscopic interaction rules. 
We generalize the principle of maximum entropy to account for
multi-time correlations between birds, and show that maximizing the entropy under this constraint is 
equivalent to inferring a dynamical model of social forces.
We test our dynamical inference method on synthetic data generated by
a topological Vicsek model (VM), showing 
that its inferred interaction parameters are consistently better than
the ones obtained in an equilibrium framework, especially when
the relative mobility between individuals is high.
When the interaction network is static, and the
dynamics satisfies detailed balance, our method recovers the results of the static approach
\cite{Bialek:2012p12539}, additionally allowing us to separate the contributions of 
interaction strength and noise to the alignment dynamics.

Maximum entropy distributions are the least constrained distributions that are consistent with certain selected key observables of the data.
They usually map onto equilibrium statistical mechanics problems and do not involve any assumptions about the system under study, 
besides the choice of the relevant observables, which should be selected accordingly to the fundamental symmetries
of the underlying system. They have been particularly successful in describing collective and
emergent phenomena in biological systems comprising many correlated degrees of freedom
\cite{maxent}. 
When considering flocks,  where polar order is present,
a natural choice of observables to be constrained by the data are the
equal time pairwise correlation functions between birds orientations: $\<s_i s_j\>$, where
$s_i$ is a $d$-dimensional unit vector denoting the flight direction of
bird $i$, with $i=1,\ldots,N$. (Throughout the paper inner products over the physical space are implicit.)
These correlations were found to exhibit scale-free behavior in natural flocks \cite{Cavagna:2010p7977}, and characterize the collective nature of flocking.  
The maximum entropy distribution $P({\bf s})$ for the orientations can then be computed by maximizing the entropy $S[P]=-\sum_\mathbf{s} P(\mathbf{s}) \ln P(\mathbf{s})$, while constraining the equal-time correlations to their experimental values. The result is  the stationary probability distribution for the 
equilibrium heterogeneous Heisenberg model~\cite{Bialek:2012p12539}:
\beq\label{eq:static}
P( \mathbf{s} )=\frac{1}{Z}\exp\lp \frac{1}{2}\sum_{i\neq j}J^{\rm stat}_{ij}s_i s_j \rp,
\eeq
where $\mathbf{s}$ is a shorthand for $(s_1,s_2,\ldots,s_N)$ and $Z$ a
normalization constant. The interaction parameters $J_{ij}^{\rm  stat}$ are Lagrange multipliers that need to be tuned so that the probability
distribution (\ref{eq:static})  matches the empirical correlation functions $\<s_i s_j\>$. 
Using 3D, single individual resolution data of large bird flocks, 
this class of models was shown to recapitulate quantitatively the ordering properties of real flocks \cite{Bialek:2012p12539}.

But infinitely many dynamical models
may give rise to this steady-state distribution, most of which break
detailed balance. In fact, the change of
neighborhoods causes the interaction network to vary in time, keeping the system constantly 
out of equilibrium.
Here we extend the maximum entropy framework to account for the non-equilibrium nature of flocking. 
We consider the set of entire trajectories $( \mathbf{s}^1, \mathbf{s}^2,\ldots, \mathbf{s}^T)$, where the
superscript index denotes time points separated by $\delta t$. We then
look for the distribution $P( \mathbf{s}^1,\ldots, \mathbf{s}^T )$
that maximizes the entropy while reproducing some given experimental
observables. Since we want to capture the dynamics, in addition to
equal-time correlation functions, we also constrain the correlation functions between two consecutive time points $\<s_i^{t+1}s_j^t\>$. Doing so yields  the following form of the probability distribution over {\em trajectories} (see Appendix for details):
\beq\label{eq:maxentdyn}
P( \mathbf{s}^1,\ldots, \mathbf{s}^T )=\frac{1}{\hat Z}\exp\lp - \mathcal{A}
\rp,
\eeq
where $\hat Z$ is a normalization factor, and the ``effective action''
(or minus log-likelihood) reads: 
\beq
\mathcal{A}=-\frac{1}{2}\sum_t \sum_{i\neq j} \lp J^{(1)}_{ij;t} s_i^t s_j^t +
J^{(2)}_{ij;t} s_i^{t+1} s_j^t\rp.
\eeq
There now are two sets of time-dependent coupling parameters, for synchronous and
consecutive times. We note that the probability (Eq.~\ref{eq:maxentdyn}) corresponds to Markovian dynamics;  non-Markov forms are possible if constraining more complex multi-time observables.

When flight orientations are highly polarized (as in the case of
starling flocks \cite{Cavagna:2010p7977}), one
can use the spin-wave (SW) approximation \cite{Dyson:1956p12932} to explicitly rewrite the action as
a sum of Markov terms which are quadratic in the spin-wave variables.
Specifically, we denote
$s_i\!\!=\!\!\pi_i \!+\! n\sqrt{1-(\pi_i)^2}$, where $n$ is
an abitrary unit vector close to the average flight direction of the
flock, and $\pi_i$ is the perpendicular component of the orientation,
$\pi_i n \!\!=\!\! 0$.
(When there is no ambiguity we drop the time superscript.)
When the flock is highly polarized,
we have $\pi_i^2\! \ll \!1$, and we may expand at small $\pi_i$. 
The action may then be written as a sum of terms corresponding to the transition probabilities 
$P(\mbox{\boldmath$\pi$}' | \mbox{\boldmath$\pi$})$ between successive
time points (see Appendix for technical details):
$\mathcal{A}+\ln\hat Z=-\ln P(\mathbf{s^1})+\sum_t \mathcal{L}_t$,
with:
\beq\label{eq:lagrange}
\begin{split}
\mathcal{L}_t( \mbox{\boldmath$\pi$}^{t+1}, \mbox{\boldmath$\pi$}^t)\equiv& - \log P(\mbox{\boldmath$\pi$}^{t+1}| \mbox{\boldmath$\pi$}^t)
=-\frac{d-1}{2}\ln{\lp \frac{\det \mathbf{A}_t}{(2\pi)^N}\rp}\\
&{ +\frac{1}{2} \lp \mbox{\boldmath$\pi$}^{t+1} -
\mathbf{M}_t \mbox{\boldmath$\pi$}^t\rp^{\dagger}\mathbf{A}_t \lp \mbox{\boldmath$\pi$}^{t+1} -
\mathbf{M}_t \mbox{\boldmath$\pi$}^t\rp},
\end{split}
\eeq
where $\mathcal{L}_t$ is formally equivalent to a Lagrangian density. 
In \eqref{eq:lagrange}
we have defined:
$\mathbf{M}_t\!\!=\!\! \mathbf{A}_t^{-1} \mathbf{J}^{(2)}_t/2$ with
$A_{ij;t}\!\!=\!\!-K_{ij;t} \!+\! \delta_{ij}\sum_k K_{ik;t}
\!+\! \delta_{ij}\sum_{k}J^{(2)}_{ik;t}/2$, where $ \mathbf{K}_t$ is a calculation intermediate obtained
by a descending recursion enforcing normalization at
each time step: $ \mathbf{K}_{t-1}= \mathbf{J}^{(1)}_t+{ \mathbf{J}^{(2)\dagger}_t} \mathbf{A}_t^{-1} \mathbf{J}^{(2)}_t/4$.

The Gaussian form of the transition probabilities Eq.~\eqref{eq:lagrange}, 
corresponds to a spin-wave dynamics described by the following
stochastic equation:
\beq\label{eq:vicsek}
\pi_i^{t+1}=\sum_j M_{ij;t} \pi_j^t + \epsilon_i^t,
\eeq
with $ {\mathbf{\epsilon}}^t$ being a random, isotropic Gaussian noise perpendicular to $n$,
of zero mean and covariance: $
\<{ \epsilon}^{t} ( { \epsilon}^{t'})^{\dagger}\> \!\!=\!\!2(d\!-\!1)\mathbf{A}_t^{-1}\delta_{t,t'}$,
where $\delta_{t,t'}$ is the Kronecker delta. 

Eq.~\eqref{eq:vicsek} can be interpreted as follows.
At each time, individual $i$ computes its new orientation from a weighted
average over the orientation of other individuals, including itself, at the previous time point
with weights encoded in the matrix $\mathbf{M}_t$
(one can check that, by construction, $\sum_j M_{ij}=1$).
Noise
$ {\bm \epsilon}^t$ added to this average determines the level of
error in the alignment. 
Without it, all individuals would be perfectly aligned.
This model may be viewed as the spin-wave expansion of a generalized Vicsek model \cite{Vicsek:1995p10953} with 
arbitrary weights and  noise.

Tuning the parameters to match the correlation functions is equivalent
to maximizing the likelihood, Eq.~\eqref{eq:maxentdyn} (see Appendix), or
equivalently maximizing the log-likelihood $-\sum_t \mathcal{L}_t$,
with which we will work from now on.
To maximize the likelihood with respect to the two equivalent sets of parameters $\{\mathbf{J}^{(1)}_t,\mathbf{J}^{(2)}_t\}$ 
or $\{\mathbf{M}_t, \mathbf{A}_t\}$, we would need to observe a large number of
random realizations of the same flock dynamics. This is impossible in
practice due to limited data 
compared to prohibitively large number of potential configurations of
the bird positions
that one would need to sample.

To overcome this problem, we need to introduce some additional
assumptions about the interaction network and the form of the noise
in order to simplify the parameter space and the number of observables. 
These simpifications come naturally in the Markovian description parametrized by $ \mathbf{M}_t$ and $ \mathbf{A}_t$.
From a biological standpoint, it is reasonable to assume that birds treat information from each interacting neighbor 
(the precise definition of ``neighborhood'' being left unspecified for
the moment) equally, 
while keeping memory of their own direction. Mathematically this translates into:
\beq\label{eq:param}
M_{ij}=(1-J\delta t n_i)\delta_{ij}
+ J\delta t n_{ij}\,,
\eeq
where $n_{ij}\!\!=\!\!1$ if $j$ is one of $i$'s neighbours, and 0
otherwise, and $n_i \!\!=\!\! \sum_{j}n_{ij}$ is the global number
of neighbors interacting with bird $i$.
(For ease of notation we omit the $t$ index,
even though $n_{ij}$ depends on $t$.)
 The scalar parameter $J$ now measures the alignemnt interaction strength.
Errors made by different birds when trying to align with their
neighbours can be assumed to be of the same amplitude and independent of each other, so that
noise is uncorrelated and $\mathbf{A}$ is proportional to the
identity, $A_{ij} \!\!=\!\! [1/(2\delta tT)]\delta_{ij}$.
Here $T$ is a squared noise amplitude (the out-of-equilibrium equivalent of a temperature) that
sets the level of disorder in the system. The scaling in $\delta t$ ensures 
a well-defined continuous limit when $\delta t \! \to \! 0$, described by a Langevin
equation.

We can reconcile this dynamical description with the static inference \cite{Bialek:2012p12539}  in the special case of equilibrium dynamics, which is realized when $n_{ij}$ is symmetric and constant in time.
In this case, the spins can be described for $\delta t\to 0$ by a stationary distribution with the same form as in Eq.~(\ref{eq:static}) 
and the steady-state couplings take the simple equilibrium value \cite{Bialek:2012p12539}, $J^{\rm stat}_{ij} \!\!=\!\! (J/T)
n_{ij}$ (see Appendix). 

Taking the specific form of $ \mathbf{M}_t$ and $ \mathbf{A}_t$ 
above for a given network of neighbours, 
we obtain a formula for $\mathcal{L}_t$ that only depends on 
two parameters, the interaction strength $J$ and the ``effective temperature''
$T$
\bea
&&\mathcal{L}_t=\frac{d-1}{2}\ln (2T\delta t)-2Jn_c\delta t \left [\tilde{C}_{s}-C_{\rm int}-\tilde{G}_s+G_{\rm int}\right ] \nonumber\\
&&+(Jn_c\delta t)^2\left [{\hat C}_s-2{\tilde C}_{\rm int}+C_{\rm int}'\right ]+C_s^1+C_s-2G_s \ ,
\label{eq:likelihood}
\eea
with $n_c=(1/N)\sum_i n_i$. Also, the number of independent observables appearing in $\mathcal{L}_t$ is drastically
reduced, to a handful of empirical integrated pair correlation functions defined
in Table~\ref{table}.  These correlations can be evaluated over pairs of consecutive configurations, or averaged over the entire sequence if we work with time-independent parameters  and steady state dynamics. 
 
 Maximizing the log-likelihood with respect to $J$ and $T$, $\partial \mathcal{L}_t/\partial T \!\!=\!\! 0$ and $\partial \mathcal{L}_t/\partial
J \!\!=\!\! 0$, yields simple analytical expressions for the parameters as a
function of the empirical correlation functions:
\bea
    J&=&\frac{1}{n_c}\frac{\Omega + (d-1)T_0}
    {C'_{\rm int}+\hat{C}_s-2\tilde{C}_{\rm
        int}},\label{eq:inference}\\
T &=& T_0+\frac{C_s^1-C_s}{2(d-1)\delta t} - \frac{J\,n_c\delta
  t}{2(d-1)}\left(\frac{\tilde C_s-\tilde G_s}{\delta t}+\Omega\right),\label{eq:T}
\eea
where
\beq
T_0=\frac{{C}_s-{G}_s}{\delta t(d-1)}, \quad \Omega=\frac{G_{\rm int}-C_{\rm int}}{\delta t}.
\eeq
The leading-order temperature $T_0$ is the derivative of a self-correlation
function, and obeys the standard fluctuation-dissipation relationship
found in equilibrium dynamics. The term $\Omega$ is
related to the dynamics of the network. In particular, at
steady state $dC_{\rm int}/dt=0$ implies $\Omega \propto \sum_{ij}
\pi_i\pi_j  dn_{ij}/dt$.

\begin{table}
\begin{center}
\begin{tabular}{|l|l||l|l|}
\hline
$C_s^1$ & $(1/N)\sum_i (\pi_i^{t+1})^2$ & $C_{\rm int}$& $(1/Nn_c)\sum_{ij}n_{ij}\pi_i^t\pi_j^t$\\
\hline
$C_s$& $(1/N)\sum_i (\pi_i^t)^2$ & $C'_{\rm
  int}$&$ (1/Nn_c^2)\sum_{ijk}n_{ij}n_{ik}\pi_j^t\pi_k^t$\\
\hline
$G_s$
&$(1/N)\sum_i \pi_i^{t+1}\pi_i^t$ & $G_{\rm int}$
&$(1/Nn_c)\sum_{ij}n_{ij}\pi_i^{t+1}\pi_j^t$\\
\hline
$\tilde C_s$
&$(1/Nn_c)\sum_i n_i (\pi_i^{t})^2$
& $\hat C_{s}$
&$(1/Nn_c^2)\sum_{ij}(n_{i} \pi_i^{t})^2$\\
\hline
$\tilde G_s$
&$(1/Nn_c)\sum_i n_i \pi_i^{t+1}\pi_i^t$
& $\tilde C_{\rm int}$
&$(1/Nn_c^2)\sum_{ij}n_{i}n_{ij}\pi_i^{t}\pi_j^t$\\
\hline
\end{tabular}
\caption{\label{table}Empirical correlation
  functions used in the text.}
\end{center}
\end{table}

In order to apply Eqs. (\ref{eq:inference})-(\ref{eq:T}) to data, one still needs to specify the neighboring matrix $n_{ij}$. 
In absence of prior information, the simplest possibility is to assume that each bird interacts with the first $n_c$ neighbors \cite{Bialek:2012p12539}. An alternative choice would be to define neighbors according to a metric rule, each bird interacting with neighbors within a given distance $r_c$. In both cases an extra parameter is introduced, either the `topological' interaction range $n_c$ or the metric range $r_c$, that can also be inferred by likelihood maximization. Another scheme is to define neighbors through a Voronoi  tassellation \cite{Voronoi}, as in the Topological VM \cite{Ginelli:2010p12925}. Likelihoods between different neighborhood definitions may also be compared to find the one closest to optimality.

We tested our dynamical inference method on synthetic data generated from a
slight generalization of the Topological VM on a two dimensional torus
of linear size $L=32$ with $N=1024$ particles:
\bea
\label{VM1}
\theta_i^{t+\delta t} &=&\mathrm{Arg}[ s_i^t + J_V\delta t \sum_j
n_{ij}s_j^t ]
+\sqrt{\delta t}\,\xi_i^t\\
\label{VM2}
r_i^{t+\delta t}&=&r_i^t+v_0\,\delta t \, s_i^{t+\delta t}
\eea
where $s_i \!\!=\!\!(\cos \theta_i , \sin \theta_i)$, and $\mathrm{Arg}(s)$ is the
angle of vector $s$.
The delta-correlated
angular noise $\xi_i^t$ is uniformly distributed in
$[-\eta\pi,+\eta\pi]$, corresponding
to an effective temperature $T_V \!\!=\!\!(\eta \pi^2)/6$ for $\delta
t \!\to \! 0$.
The Voronoi adjacency matrix $n_{ij}$ has a non-uniform
degree $n_i$, of mean $n_V\!\!=\!\!6$. A spin-wave expansion of 
Eq. (\ref{VM1}) leads to an expression
of the form of (\ref{eq:vicsek})-(\ref{eq:param}),
with $J\approx J_V/(1+J_Vn_V\delta t)$ (see Appendix).
The degree of neighbor mixing is characterized by a single mixing
parameter $ \mu \!\!=\!\!\langle 1/(Nn_c)\sum_{i j} |d{n}_{ij}/dt|\rangle$, 
which quantifies how fast birds exchange neighbors. We performed
simulations with time step $\delta t=0.01$ in three regimes with
slow, medium and fast neighbor mixing ($\mu=0.18,0.35,0.76$,
$v_0=0.5,1.0,2.0$, $J_V=1.0,1.0,0.1$ and $\eta=0.3,0.2,0.12$ respectively),
all of which display the same level of polarization,
$N^{-1}\Vert\sum_i s_i \Vert \approx 0.97$.

\begin{figure}
\noindent\includegraphics[width=.79\linewidth]{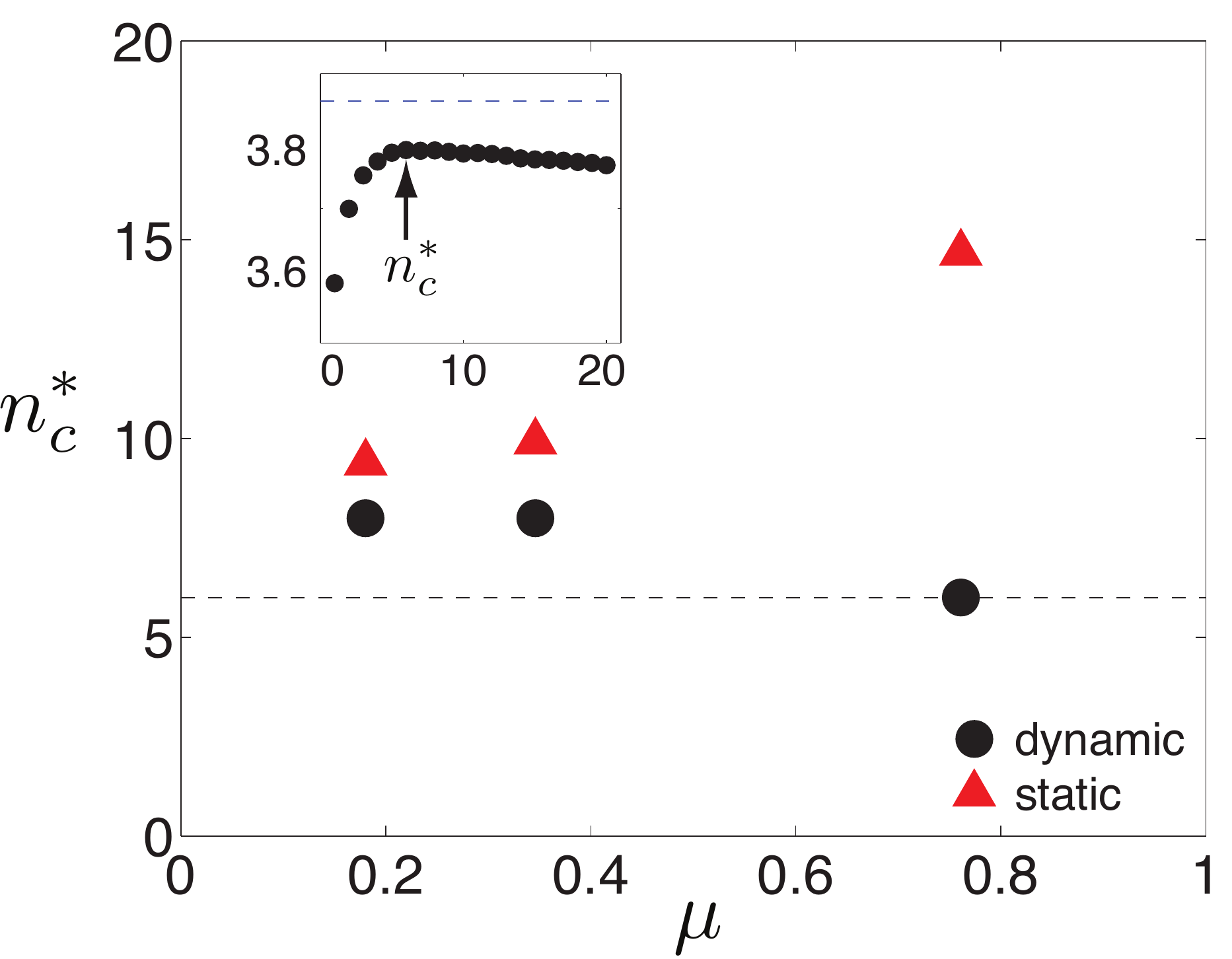}
\caption{
\label{fig:all}
(Color online.) Comparison between dynamical and 
static inference. Data was generated using Voronoi neighborhood. 
The inference was performed by using either a Voronoi rule
or a nearest-neighbor (NN) topological rule, parametrized by the
number $n_c$ of interacting neighbors.
Main panel: The inferred number of interacting neighbors $n_c^*$
is shown as a function of the mixing rate $\mu$. Circles: dynamical inference; triangles: static inference. 
The dashed lines marks the real
average value, $n_V=6$. Static inference badly overestimates
the number of interacting neighbors at large mixing, while dynamical inference does a much better job. 
Inset: Dynamical normalized log-likelihood $-\mathcal{L}_t/N$ as
a function of $n_c$ for the NN topological rule (circles). The maximum of this function
gives the NN value of $n_c^*$ reported in the main panel.
The Voronoi likelihood (dashed line) is larger than the NN one, revealing that 
Voronoi was the actual generating rule. Data are for high
mixing.
}
\end{figure}

We then  applied the inference procedure described in Eqs.~(\ref{eq:inference})-(\ref{eq:T}) to the 
synthetic dataset generated by the simulations. In the inference we tried the 
choices for $n_{ij}$ discussed above: the $n_c$ nearest-neighbor (NN)
topological rule, the metric rule where $n_{ij}=1$ within a metric
range $r_c$ (and $0$ outside),  and the
Voronoi rule (actually used to generate the data).
Correlation functions were averaged over $10^3$ different
configurations in the stationary state, sampled from a single run at
100 time unit intervals, ensuring independent sampling.

The likelihood as a function of $n_c$ can be computed with the NN rule using Eqs.~(\ref{eq:inference})(\ref{eq:T}) and (\ref{eq:likelihood}).
The result is shown in the inset of Fig.~\ref{fig:all} for the high mixing regime. Its maximum $n_c^*$ corresponds to the most likely interaction
range, from which the optimal $J^*$ and
$T^*$ are computed via
Eqs.~(\ref{eq:inference})-(\ref{eq:T}).
Fig.~\ref{fig:all}  shows that the new dynamical procedure systematically
outperforms the static approach described in \cite{Bialek:2012p12539} in predicting the mean
interaction range $n_c$. 
The error made by the static inference is larger when neighbor mixing
is higher and the dynamics is strongly
out-of-equilibrium. That is because in the high-mixing case, the {\em
  effective} number of interacting neighbors,
as inferred by the static approach, includes neighbors visited in the recent
past in addition to the current ones, and thus is larger than the true
$n_c$.
Overall, the dynamical inference based on NN interactions performs
reasonably well, considering that the model used for the inference
incorrectly assumes a constant $n_c$.
Not surprisingly, the log-likelihood computed with the (correct) Voronoi topology
is larger than with the (incorrect) NN one.
The temperature $T$ is well inferred in both cases  (8\% error),
while the alignment strength $J$ is well recovered when assuming Voronoi
neighbors (3\% error), and approximately with a NN topology (20\%
error).

\begin{figure}
\noindent\includegraphics[width=.79\linewidth]{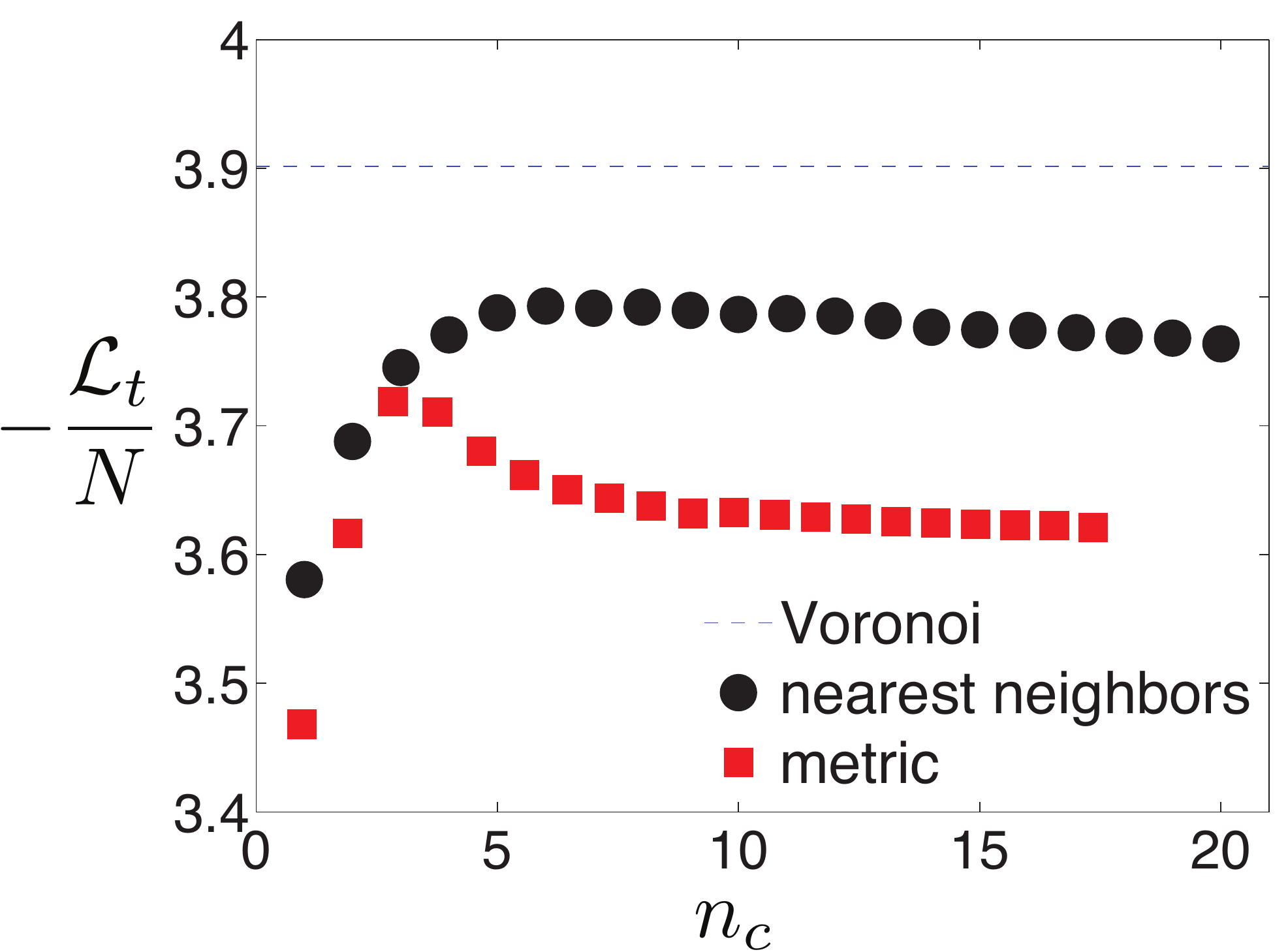}
\noindent\includegraphics[width=.79\linewidth]{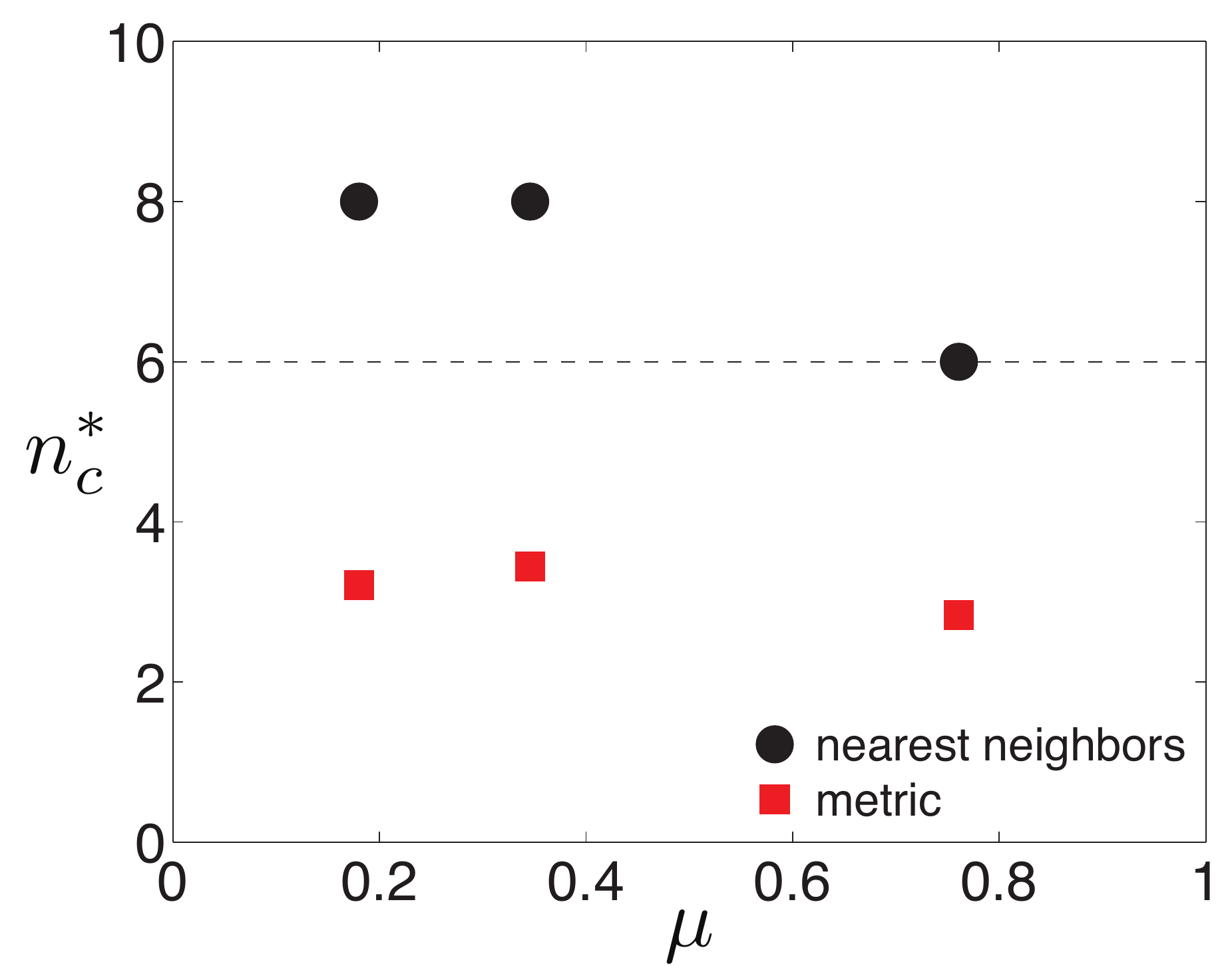}
\caption{
\label{fig:SI}
Upper: comparison of the normalized log-likelihood for the
nearest-neighbor and metric rules, as a function of $n_c$. For
the metric case, for increasing values of $r_c$, the empirical
$n_c=(1/N)\sum_i n_i$ is shown. The dashed line corresponds to the
log-likelihood calculated with the Voronoi rule. Lower: Inferred
interaction range
$n_c^*$ for the nearest-neighbor and metric cases, as a function of
the mixing parameter $\mu$.
}
\end{figure}

Performing the dynamical inference using a metric rule
gives significantly worse results, giving $n_c\sim 3$ (see figure ~\ref{fig:SI}), 
a factor $2$ smaller than the correct value. Hence the dynamical 
method not only gives us the correct interaction parameters, but 
also distinguishes the rule used to build the interaction network. 
The method exploits the different ways in which 
spatial density fluctuations translate into fluctuations in the number
of neighbors.
In the Voronoi network (the generating one), the number of 
neighbors $n_i$ of each point fluctuates weakly around its mean 
value of 6. In the NN case, $n_i$ does not fluctuate at all,
whereas with the metric rule $n_i$ exhibits {\it very} large fluctuations, 
directly linked to the VM giant density fluctuations \cite{Ginelli:2010p12925}.
The large fluctuations of $n_i$ make the correlation functions
of Table \ref{table}
very different from their correct (Voronoi) values.

In summary, we have derived a dynamical maximum entropy method to
infer the alignment dynamics of highly-ordered animal groups from just
two consecutive snapshots.
Tests on synthetic data confirm the validity of our method.
Our approach is very general
and makes minimal, symmetry-based assumptions on the structure of the
dynamics under investigation, 
alternative to other inference methods
\cite{birdblock}. 
Related approaches have been proposed in the context of Ising
spins or spiking neurons
\cite{morespins}; however, in that case 
it is hard to relate a simple interaction form of the Markovian transition probabilities
to a principle of maximum entropy.
Our work emphasizes the need
for a dynamical inference approach to
out-of-equilibrium active matter systems, especially when there is no {\em a
  priori} knowledge of the timescales in the system, which is usually the case
when dealing with experimental data.

Our approach is applicable to many 
systems where collective motion is observed, including
moving animal groups \cite{Parrish}, bacterial colonies
\cite{twoblock}, motility assays \cite{Sumino},
collective motion of epithelial cells \cite{Sepulveda:2013p13003}, 
or vibrated polar disks 
\cite{twoblock2}. 
Throughout this work we have assumed that
$\delta t$ is equal to (or smaller than) the real update time lag,
namely the biological timescale. This may not be true for some
datasets, as the sampling time of the experimental equipment is likely
to be larger than the neural update time actually used by
animals. This is certainly the case for the starling data of
\cite{Bialek:2012p12539}. When this happens, the experimental time
series is a coarse-grained version of the real dynamics, so that the
present method would probably provide a time-renormalized value of the
interaction parameters. It would therefore be important to generalize
our equations to deal with this issue.
Other generalizations include the analysis of
other symmetries than the polar one (as in systems with nematic order \cite{rods}), or the
extension to second-order dynamics describing systems
characterized by linear, not diffusive, dispersion relations
\cite{Attanasi:2013p12927}.

Acknowledgements. We thank Martin Weigt for helpful discussions.
I.G. was supported by grants IIT--Seed Artswarm, ERC--StG n.257126. 
A.C was supported by grant US-AFOSR FA95501010250 (through the University of Maryland).
FG acknowledges support from grants EPSRC First Grant EP/K018450/1 and
MC Career Integration Grant PCIG13-GA-2013-618399.
Work in Paris was supported by grant ERC–StG n. 306312.

\bibliography{thierry_clean,byhand}

\begin{thebibliography}{10}
\makeatletter
\providecommand \@ifxundefined [1]{%
 \ifx #1\undefined \expandafter \@firstoftwo
 \else \expandafter \@secondoftwo
\fi
}%
\providecommand \@ifnum [1]{%
 \ifnum #1\expandafter \@firstoftwo
 \else \expandafter \@secondoftwo
\fi
}%
\providecommand \enquote [1]{``#1''}%
\providecommand \bibnamefont  [1]{#1}%
\providecommand \bibfnamefont [1]{#1}%
\providecommand \citenamefont [1]{#1}%
\providecommand\href[0]{\@sanitize\@href}%
\providecommand\@href[1]{\endgroup\@@startlink{#1}\endgroup\@@href}%
\providecommand\@@href[1]{#1\@@endlink}%
\providecommand \@sanitize [0]{\begingroup\catcode`\&12\catcode`\#12\relax}%
\@ifxundefined \pdfoutput {\@firstoftwo}{%
 \@ifnum{\z@=\pdfoutput}{\@firstoftwo}{\@secondoftwo}%
}{%
 \providecommand\@@startlink[1]{\leavevmode\special{html:<a href="#1">}}%
 \providecommand\@@endlink[0]{\special{html:</a>}}%
}{%
 \providecommand\@@startlink[1]{%
  \leavevmode
  \pdfstartlink
   attr{/Border[0 0 1 ]/H/I/C[0 1 1]}%
   user{/Subtype/Link/A<</Type/Action/S/URI/URI(#1)>>}%
  \relax
 }%
 \providecommand\@@endlink[0]{\pdfendlink}%
}%
\providecommand \url  [0]{\begingroup\@sanitize \@url }%
\providecommand \@url [1]{\endgroup\@href {#1}{\urlprefix}}%
\providecommand \urlprefix [0]{URL }%
\providecommand \Eprint[0]{\href }%
\@ifxundefined \urlstyle {%
  \providecommand \doi [1]{doi:\discretionary{}{}{}#1}%
}{%
  \providecommand \doi [0]{doi:\discretionary{}{}{}\begingroup
  \urlstyle{rm}\Url }%
}%
\providecommand \doibase [0]{http://dx.doi.org/}%
\providecommand \Doi[1]{\href{\doibase#1}}%
\providecommand \bibAnnote [3]{%
  \BibitemShut{#1}%
  \begin{quotation}\noindent
    \textsc{Key:}\ #2\\\textsc{Annotation:}\ #3%
  \end{quotation}%
}%
\providecommand \bibAnnoteFile [2]{%
  \IfFileExists{#2}{\bibAnnote {#1} {#2} {\input{#2}}}{}%
}%
\providecommand \typeout [0]{\immediate \write \m@ne }%
\providecommand \selectlanguage [0]{\@gobble}%
\providecommand \bibinfo [0]{\@secondoftwo}%
\providecommand \bibfield [0]{\@secondoftwo}%
\providecommand \translation [1]{[#1]}%
\providecommand \BibitemOpen[0]{}%
\providecommand \bibitemStop [0]{}%
\providecommand \bibitemNoStop [0]{.\EOS\space}%
\providecommand \EOS [0]{\spacefactor3000\relax}%
\providecommand \BibitemShut [1]{\csname bibitem#1\endcsname}%
\bibitem{Vicsek:1995p10953}%
  \BibitemOpen
  \bibfield{author}{%
  \bibinfo {author} {\bibfnamefont{T.}~\bibnamefont{Vicsek}}, \bibinfo {author}
  {\bibfnamefont{A.}~\bibnamefont{Czir{\'o}k}}, \bibinfo {author}
  {\bibfnamefont{E.}~\bibnamefont{Ben-Jacob}}, \bibinfo {author}
  {\bibfnamefont{I.}~\bibnamefont{Cohen}},\ and\ \bibinfo {author}
  {\bibfnamefont{O.}~\bibnamefont{Shochet}},\ }%
  \bibfield{journal}{%
  \bibinfo {journal} {PRL}\ }%
  \textbf{\bibinfo {volume} {75}},\ \bibinfo {pages} {1226} (\bibinfo {month}
  {Aug}\ \bibinfo {year} {1995})%
  \bibAnnoteFile{NoStop}{Vicsek:1995p10953}%
\bibitem{block1}
  \BibitemOpen
  \bibfield{author}{%
  \bibinfo {author} {\bibfnamefont{Y.}~\bibnamefont{Tu}}, \bibinfo {author}
  {\bibfnamefont{J.}~\bibnamefont{Toner}},\ and\ \bibinfo {author}
  {\bibfnamefont{M.}~\bibnamefont{Ulm}},\ }%
  \bibfield{journal}{%
  \Doi{10.1103/PhysRevLett.80.4819}{\bibinfo {journal} {PRL}}\ }%
  \textbf{\bibinfo {volume} {80}},\ \bibinfo {pages} {4819} (
 \bibinfo {year} {1998});\ %
  \bibAnnoteFile{NoStop}{Tu:1998p12906}%
  \bibfield{author}{%
  \bibinfo {author} {\bibfnamefont{G.}~\bibnamefont{Gr{\'e}goire}}\ and\
  \bibinfo {author} {\bibfnamefont{H.}~\bibnamefont{Chat{\'e}}},\ }%
  \bibfield{journal}{%
  \bibinfo {journal} {PRL}\ }%
  \textbf{\bibinfo {volume} {92}},\ \bibinfo {pages} {025702} (
 \bibinfo {year} {2004});\ %
  \bibAnnoteFile{NoStop}{Gregoire:2004p12912}%
  \bibfield{author}{%
  \bibinfo {author} {\bibfnamefont{E.}~\bibnamefont{Bertin}}, \bibinfo {author}
  {\bibfnamefont{M.}~\bibnamefont{Droz}},\ and\ \bibinfo {author}
  {\bibfnamefont{G.}~\bibnamefont{Gr\'{e}goire}},\ }%
  \bibfield{journal}{%
  \bibinfo {journal} {PRE}\ }%
  \textbf{\bibinfo {volume} {74}},\ \bibinfo {pages} {022101} (\bibinfo {year}
  {2006});\ %
  \bibAnnoteFile{NoStop}{Bertin1}%
  \bibfield{author}{%
  \bibinfo {author} {\bibfnamefont{E.}~\bibnamefont{Bertin}}, \bibinfo {author}
  {\bibfnamefont{M.}~\bibnamefont{Droz}},\ and\ \bibinfo {author}
  {\bibfnamefont{G.}~\bibnamefont{Gr\'{e}goire}},\ }%
  \bibfield{journal}{%
  \bibinfo {journal} {J. Phys. A}\ }%
  \textbf{\bibinfo {volume} {42}},\ \bibinfo {pages} {445001} (\bibinfo {year}
  {2009});\ %
  \bibAnnoteFile{NoStop}{Bertin2}%
  \bibfield{author}{%
  \bibinfo {author} {\bibfnamefont{T.}~\bibnamefont{Ihle}},\ }%
  \bibfield{journal}{%
  \bibinfo {journal} {PRE}\ }%
  \textbf{\bibinfo {volume} {83}},\ \bibinfo {pages} {030901} (\bibinfo {year}
  {2011});\ %
  \bibAnnoteFile{NoStop}{Ihle}%
  \bibfield{author}{%
  \bibinfo {author} {\bibfnamefont{H.}~\bibnamefont{Chat{\'e}}}, \bibinfo
  {author} {\bibfnamefont{F.}~\bibnamefont{Ginelli}}, \bibinfo {author}
  {\bibfnamefont{G.}~\bibnamefont{Gr{\'e}goire}},\ and\ \bibinfo {author}
  {\bibfnamefont{F.}~\bibnamefont{Raynaud}},\ }%
  \bibfield{journal}{%
  {PRE}\ }%
  \textbf{\bibinfo {volume} {77}},\ \bibinfo {pages} {046113} (
\bibinfo {year} {2008});\ %
  \bibAnnoteFile{NoStop}{Chate:2008p12924}%
  \bibfield{author}{%
  \bibinfo {author} {\bibfnamefont{P.}~\bibnamefont{Szab{\'o}}}, \bibinfo
  {author} {\bibfnamefont{M.}~\bibnamefont{Nagy}},\ and\ \bibinfo {author}
  {\bibfnamefont{T.}~\bibnamefont{Vicsek}},\ }%
  \bibfield{journal}{%
  \bibinfo {journal} {PRE }\ }%
  \textbf{\bibinfo {volume} {79}},\ \bibinfo {pages} {021908} (
 \bibinfo {year} {2009});\ %
  \bibAnnoteFile{NoStop}{Szabo:2009p12929}%
  \bibfield{author}{%
  \bibinfo {author} {\bibfnamefont{A.}~\bibnamefont{Peshkov}}, \bibinfo
  {author} {\bibfnamefont{S.}~\bibnamefont{Ngo}}, \bibinfo {author}
  {\bibfnamefont{E.}~\bibnamefont{Bertin}}, \bibinfo {author}
  {\bibfnamefont{H.}~\bibnamefont{Chat\'{e}}},\ and\ \bibinfo {author}
  {\bibfnamefont{F.}~\bibnamefont{Ginelli}},\ }%
  \bibfield{journal}{%
  \bibinfo {journal} {PRL}\ }%
  \textbf{\bibinfo {volume} {109}},\ \bibinfo {pages} {098101} (\bibinfo {year}
  {2012});\ %
  \bibAnnoteFile{NoStop}{Bertin3}%
  \bibfield{author}{%
  \bibinfo {author} {\bibfnamefont{J.}~\bibnamefont{Toner}},\ }%
  \bibfield{journal}{%
  \bibinfo {journal} {PRE}\ }%
  \textbf{\bibinfo {volume} {86}},\ \bibinfo {pages} {031918} (
\bibinfo {year} {2012});\ %
  \bibAnnoteFile{NoStop}{Toner:2012p12923}%
  \bibfield{author}{%
  \bibinfo {author} {\bibfnamefont{S.}~\bibnamefont{Ramaswamy}},\ }%
  \bibfield{journal}{%
  \bibinfo {journal} {Annu. Rev. Condens. Matter Phys.}\ }%
  \textbf{\bibinfo {volume} {1}},\ \bibinfo {pages} {323} (\bibinfo {year}
  {2010})%
  \bibAnnoteFile{NoStop}{Review}%
\bibitem{TonerTu}%
  \BibitemOpen
  \bibfield{author}{%
  \bibinfo {author} {\bibfnamefont{J.}~\bibnamefont{Toner}}\ and\ \bibinfo
  {author} {\bibfnamefont{Y.}~\bibnamefont{Tu}},\ }%
  \bibfield{journal}{%
  \bibinfo {journal} {PRL}\ }%
  \textbf{\bibinfo {volume} {75}},\ \bibinfo {pages} {4326} (
 \bibinfo {year} {1995});\ %
  \bibAnnoteFile{NoStop}{Toner:1995p11446}%
  \bibfield{author}{%
  \bibinfo {author} {\bibfnamefont{J.}~\bibnamefont{Toner}}\ and\ \bibinfo
  {author} {\bibfnamefont{Y.}~\bibnamefont{Tu}},\ }%
  \bibfield{journal}{%
  \bibinfo {journal} {PRE}\ }%
  \textbf{\bibinfo {volume} {58}},\ \bibinfo {pages} {4828} (
 \bibinfo {year} {1998})%
  \bibAnnoteFile{NoStop}{Toner:1998p11445}%
\bibitem{Ginelli:2010p12925}%
  \BibitemOpen
  \bibfield{author}{%
  \bibinfo {author} {\bibfnamefont{F.}~\bibnamefont{Ginelli}}\ and\ \bibinfo
  {author} {\bibfnamefont{H.}~\bibnamefont{Chat{\'e}}},\ }%
  \bibfield{journal}{%
  \bibinfo {journal} {PRL}\ }%
  \textbf{\bibinfo {volume} {105}},\ \bibinfo {pages} {168103} (
\bibinfo {year} {2010})%
  \bibAnnoteFile{NoStop}{Ginelli:2010p12925}%
 \bibitem{Cavagnafriends}%
  \BibitemOpen
  \bibfield{author}{%
  \bibinfo {author} {\bibfnamefont{A.}~\bibnamefont{Cavagna}}, \bibinfo
  {author} {\bibfnamefont{I.}~\bibnamefont{Giardina}}, \bibinfo {author}
  {\bibfnamefont{A.}~\bibnamefont{Orlandi}}, \bibinfo {author}
  {\bibfnamefont{G.}~\bibnamefont{Parisi}}, \bibinfo {author}
  {\bibfnamefont{A.}~\bibnamefont{Procaccini}}, \bibinfo {author}
  {\bibfnamefont{M.}~\bibnamefont{Viale}},\ and\ \bibinfo {author}
  {\bibfnamefont{V.}~\bibnamefont{Zdravkovic}},\ }%
  \bibfield{journal}{%
  \Doi{10.1016/j.anbehav.2008.02.002}{\bibinfo {journal} {Anim Behav}}\ }%
  \textbf{\bibinfo {volume} {76}},\ \bibinfo {pages} {217} (
  \bibinfo {year} {2008});\ %
  \bibAnnoteFile{NoStop}{Cavagna:2008p7911}%
  \bibfield{author}{%
  \bibinfo {author} {\bibfnamefont{A.}~\bibnamefont{Cavagna}}, \bibinfo
  {author} {\bibfnamefont{I.}~\bibnamefont{Giardina}}, \bibinfo {author}
  {\bibfnamefont{A.}~\bibnamefont{Orlandi}}, \bibinfo {author}
  {\bibfnamefont{G.}~\bibnamefont{Parisi}},\ and\ \bibinfo {author}
  {\bibfnamefont{A.}~\bibnamefont{Procaccini}},\ }%
  \bibfield{journal}{%
  \Doi{10.1016/j.anbehav.2008.02.003}{ {Anim Behav}}\ }%
  \textbf{\bibinfo {volume} {76}},\ \bibinfo {pages} {237} (
   \bibinfo {year} {2008});\ %
  \bibAnnoteFile{NoStop}{Cavagna:2008p7912}%
  \bibfield{author}{%
  \bibinfo {author} {\bibfnamefont{M.}~\bibnamefont{Ballerini et al.}},} 
  \bibfield{journal}{%
  \Doi{10.1016/j.anbehav.2008.02.004}{\bibinfo {journal} {Anim Behav}}\ }%
  \textbf{\bibinfo {volume} {76}},\ \bibinfo {pages} {201} (
   \bibinfo {year} {2008})%
  \bibAnnoteFile{NoStop}{Ballerini:2008p7908}%
\bibitem{Ballerini:2008p679}%
  \BibitemOpen
  \bibfield{author}{%
  \bibinfo {author} {\bibfnamefont{M.}~\bibnamefont{Ballerini et al.}},} 
 \bibfield{journal}{%
  \Doi{10.1073/pnas.0711437105}{\bibinfo {journal} {PNAS}}\
  }%
  \textbf{\bibinfo {volume} {105}},\ \bibinfo {pages} {1232} (
   \bibinfo {year} {2008})%
  \bibAnnoteFile{NoStop}{Ballerini:2008p679}%
\bibitem{Cavagna:2010p7977}%
  \BibitemOpen
  \bibfield{author}{%
  \bibinfo {author} {\bibfnamefont{A.}~\bibnamefont{Cavagna}}, \bibinfo
  {author} {\bibfnamefont{A.}~\bibnamefont{Cimarelli}}, \bibinfo {author}
  {\bibfnamefont{I.}~\bibnamefont{Giardina}}, \bibinfo {author}
  {\bibfnamefont{G.}~\bibnamefont{Parisi}}, \bibinfo {author}
  {\bibfnamefont{R.}~\bibnamefont{Santagati}}, \bibinfo {author}
  {\bibfnamefont{F.}~\bibnamefont{Stefanini}},\ and\ \bibinfo {author}
  {\bibfnamefont{M.}~\bibnamefont{Viale}},\ }%
  \bibfield{journal}{%
  \Doi{10.1073/pnas.1005766107}{\bibinfo {journal} {PNAS}}\
  }%
  \textbf{\bibinfo {volume} {107}},\ \bibinfo {pages} {11865} (
 \bibinfo {year} {2010})%
  \bibAnnoteFile{NoStop}{Cavagna:2010p7977}%
\bibitem{Bialek:2012p12539}%
  \BibitemOpen
  \bibfield{author}{%
  \bibinfo {author} {\bibfnamefont{W.}~\bibnamefont{Bialek}}, \bibinfo {author}
  {\bibfnamefont{A.}~\bibnamefont{Cavagna}}, \bibinfo {author}
  {\bibfnamefont{I.}~\bibnamefont{Giardina}}, \bibinfo {author}
  {\bibfnamefont{T.}~\bibnamefont{Mora}}, \bibinfo {author}
  {\bibfnamefont{E.}~\bibnamefont{Silvestri}}, \bibinfo {author}
  {\bibfnamefont{M.}~\bibnamefont{Viale}},\ and\ \bibinfo {author}
  {\bibfnamefont{A.~M.}\ \bibnamefont{Walczak}},\ }%
  \bibfield{journal}{%
  \Doi{10.1073/pnas.1118633109}{\bibinfo {journal} {PNAS}}\
  }%
  \textbf{\bibinfo {volume} {109}},\ \bibinfo {pages} {4786} (
\bibinfo {year} {2012})%
  \bibAnnoteFile{NoStop}{Bialek:2012p12539}%
\bibitem{Bialek:2013p13047}%
  \BibitemOpen
  \bibfield{author}{%
  \bibinfo {author} {\bibfnamefont{W.}~\bibnamefont{Bialek}}, \bibinfo {author}
  {\bibfnamefont{A.}~\bibnamefont{Cavagna}}, \bibinfo {author}
  {\bibfnamefont{I.}~\bibnamefont{Giardina}}, \bibinfo {author}
  {\bibfnamefont{T.}~\bibnamefont{Mora}}, \bibinfo {author}
  {\bibfnamefont{O.}~\bibnamefont{Pohl}}, \bibinfo {author}
  {\bibfnamefont{E.}~\bibnamefont{Silvestri}}, \bibinfo {author}
  {\bibfnamefont{M.}~\bibnamefont{Viale}},\ and\ \bibinfo {author}
  {\bibfnamefont{A.}~\bibnamefont{Walczak}},\ }%
  \bibfield{journal}{%
  \bibinfo {journal} {arXiv:1307.5563v1}\ }%
(
\bibinfo
  {year} {2013})
  \bibAnnoteFile{NoStop}{Bialek:2013p13047}%
\bibitem{maxent}%
  \BibitemOpen
  \bibfield{author}{%
  \bibinfo {author} {\bibfnamefont{E.}~\bibnamefont{Schneidman}}, \bibinfo
  {author} {\bibfnamefont{M.~J.}\ \bibnamefont{Berry}}, \bibinfo {author}
  {\bibfnamefont{R.}~\bibnamefont{Segev}},\ and\ \bibinfo {author}
  {\bibfnamefont{W.}~\bibnamefont{Bialek}},\ }%
  \bibfield{journal}{%
  \Doi{10.1038/nature04701}{\bibinfo {journal} {Nature}}\ }%
  \textbf{\bibinfo {volume} {440}},\ \bibinfo {pages} {1007} (
   \bibinfo {year} {2006});\ %
  \bibAnnoteFile{NoStop}{Schneidman:2006p1273}%
  \bibfield{author}{%
  \bibinfo {author} {\bibfnamefont{J.}~\bibnamefont{Shlens}}, \bibinfo {author}
  {\bibfnamefont{G.~D.}\ \bibnamefont{Field}}, \bibinfo {author}
  {\bibfnamefont{J.~L.}\ \bibnamefont{Gauthier}}, \bibinfo {author}
  {\bibfnamefont{M.~I.}\ \bibnamefont{Grivich}}, \bibinfo {author}
  {\bibfnamefont{D.}~\bibnamefont{Petrusca}}, \bibinfo {author}
  {\bibfnamefont{A.}~\bibnamefont{Sher}}, \bibinfo {author}
  {\bibfnamefont{A.~M.}\ \bibnamefont{Litke}},\ and\ \bibinfo {author}
  {\bibfnamefont{E.~J.}\ \bibnamefont{Chichilnisky}},\ }%
  \bibfield{journal}{%
  \Doi{10.1523/JNEUROSCI.1282-06.2006}{\bibinfo {journal} {J Neurosci}}\ }%
  \textbf{\bibinfo {volume} {26}},\ \bibinfo {pages} {8254} (
  \bibinfo {year} {2006});\ %
  \bibAnnoteFile{NoStop}{Shlens:2006p1442}%
  \bibfield{author}{%
  \bibinfo {author} {\bibfnamefont{M.}~\bibnamefont{Weigt}}, \bibinfo {author}
  {\bibfnamefont{R.~A.}\ \bibnamefont{White}}, \bibinfo {author}
  {\bibfnamefont{H.}~\bibnamefont{Szurmant}}, \bibinfo {author}
  {\bibfnamefont{J.~A.}\ \bibnamefont{Hoch}},\ and\ \bibinfo {author}
  {\bibfnamefont{T.}~\bibnamefont{Hwa}},\ }%
  \bibfield{journal}{%
  \Doi{10.1073/pnas.0805923106}{\bibinfo {journal} {PNAS}}\
  }%
  \textbf{\bibinfo {volume} {106}},\ \bibinfo {pages} {67} (
   \bibinfo {year} {2009});\ %
  \bibAnnoteFile{NoStop}{Weigt:2009p3341}%
  \bibfield{author}{%
  \bibinfo {author} {\bibfnamefont{T.}~\bibnamefont{Mora}}, \bibinfo {author}
  {\bibfnamefont{A.~M.}\ \bibnamefont{Walczak}}, \bibinfo {author}
  {\bibfnamefont{W.}~\bibnamefont{Bialek}},\ and\ \bibinfo {author}
  {\bibfnamefont{C.~G.}\ \bibnamefont{Callan}},\ }%
  \bibfield{journal}{%
  \Doi{10.1073/pnas.1001705107}{\bibinfo {journal} {PNAS}}\
  }%
  \textbf{\bibinfo {volume} {107}},\ \bibinfo {pages} {5405} (
  \bibinfo {year} {2010});\ %
  \bibAnnoteFile{NoStop}{Mora:2010p5398}%
  \bibfield{author}{%
  \bibinfo {author} {\bibfnamefont{M.}~\bibnamefont{Santolini}}, \bibinfo
  {author} {\bibfnamefont{T.}~\bibnamefont{Mora}},\ and\ \bibinfo {author}
  {\bibfnamefont{V.}~\bibnamefont{Hakim}},\ }%
  \bibfield{journal}{%
  \bibinfo {journal} {arXiv:1302.4424v1}\ }%
 (
  \bibinfo
  {year} {2013});\ 
  \bibAnnoteFile{NoStop}{Santolini:2013p12933}%
  \bibfield{author}{%
  \bibinfo {author} {\bibfnamefont{T.}~\bibnamefont{Mora}}\ and\ \bibinfo
  {author} {\bibfnamefont{W.}~\bibnamefont{Bialek}},\ }%
  \bibfield{journal}{%
  \Doi{10.1007/s10955-011-0229-4}{\bibinfo {journal} {J Stat Phys}}\ }%
  \textbf{\bibinfo {volume} {144}},\ \bibinfo {pages} {268} (
 \bibinfo {year} {2011})%
  \bibAnnoteFile{NoStop}{Mora:2011p12729}%
\bibitem{Dyson:1956p12932}%
  \BibitemOpen
  \bibfield{author}{%
  \bibinfo {author} {\bibfnamefont{F.}~\bibnamefont{Dyson}},\ }%
  \bibfield{journal}{%
  \bibinfo {journal} {Physical review}\ }%
  \textbf{\bibinfo {volume} {102}},\ \bibinfo {pages} {1217} (\bibinfo {year}
  {1956})%
  \bibAnnoteFile{NoStop}{Dyson:1956p12932}%
\bibitem{Voronoi}%
  \BibitemOpen
  \bibfield{author}{%
  \bibinfo {author} {\bibfnamefont{G.}~\bibnamefont{Voronoi}},\ }%
  \bibfield{journal}{%
  \bibinfo {journal} {J Reine Angew Math}\ }%
  \textbf{\bibinfo {volume} {133}},\ \bibinfo {pages} {97} (\bibinfo {year}
  {1907})%
  \bibAnnoteFile{NoStop}{Voronoi}%
\bibitem{birdblock}%
  \BibitemOpen
  \bibfield{author}{%
  \bibinfo {author} {\bibfnamefont{J.~E.}\ \bibnamefont{Herbert-Read}},
  \bibinfo {author} {\bibfnamefont{A.}~\bibnamefont{Perna}}, \bibinfo {author}
  {\bibfnamefont{R.~P.}\ \bibnamefont{Mann}}, \bibinfo {author}
  {\bibfnamefont{T.~M.}\ \bibnamefont{Schaerf}}, \bibinfo {author}
  {\bibfnamefont{D.~J.~T.}\ \bibnamefont{Sumpter}},\ and\ \bibinfo {author}
  {\bibfnamefont{A.~J.~W.}\ \bibnamefont{Ward}},\ }%
  \bibfield{journal}{%
  \Doi{10.1073/pnas.1109355108}{\bibinfo {journal} {PNAS}}\
  }%
  \textbf{\bibinfo {volume} {108}},\ \bibinfo {pages} {18726} (
   \bibinfo {year} {2011});\ %
  \bibAnnoteFile{NoStop}{HerbertRead:2011p12915}%
  \bibfield{author}{%
  \bibinfo {author} {\bibfnamefont{Y.}~\bibnamefont{Katz}}, \bibinfo {author}
  {\bibfnamefont{K.}~\bibnamefont{Tunstr{\o}m}}, \bibinfo {author}
  {\bibfnamefont{C.~C.}\ \bibnamefont{Ioannou}}, \bibinfo {author}
  {\bibfnamefont{C.}~\bibnamefont{Huepe}},\ and\ \bibinfo {author}
  {\bibfnamefont{I.~D.}\ \bibnamefont{Couzin}},\ }%
  \bibfield{journal}{%
  \Doi{10.1073/pnas.1107583108}{\bibinfo {journal} {PNAS}}\
  }%
  \textbf{\bibinfo {volume} {108}},\ \bibinfo {pages} {18720} (
  \bibinfo {year} {2011});\ %
  \bibAnnoteFile{NoStop}{Katz:2011p12800}%
  \bibfield{author}{%
  \bibinfo {author} {\bibfnamefont{J.~Gautrais et al.}\ \bibnamefont{{\it et al.}}},\ }%
  \bibfield{journal}{%
  \bibinfo {journal} {PLoS Comp. Bio.}\ }%
  \textbf{\bibinfo {volume} {8}},\ \bibinfo {pages} {e1002678} (\bibinfo {year}
  {2012})%
  \bibAnnoteFile{NoStop}{Gautrais}%
\bibitem{morespins}%
  \BibitemOpen
  \bibfield{author}{%
  \bibinfo {author} {\bibfnamefont{O.}~\bibnamefont{Marre}}, \bibinfo {author}
  {\bibfnamefont{S.~E.}\ \bibnamefont{Boustani}}, \bibinfo {author}
  {\bibfnamefont{Y.}~\bibnamefont{Fr{\'e}gnac}},\ and\ \bibinfo {author}
  {\bibfnamefont{A.}~\bibnamefont{Destexhe}},\ }%
  \bibfield{journal}{%
  \Doi{10.1103/PhysRevLett.102.138101}{\bibinfo {journal} {PRL}}\ }%
  \textbf{\bibinfo {volume} {102}},\ \bibinfo {pages} {138101} (
   \bibinfo {year} {2009});\ %
  \bibAnnoteFile{NoStop}{Marre:2009p13087}%
  \bibfield{author}{%
  \bibinfo {author} {\bibfnamefont{Y.}~\bibnamefont{Roudi}}\ and\ \bibinfo
  {author} {\bibfnamefont{J.}~\bibnamefont{Hertz}},\ }%
  \bibfield{journal}{%
  \Doi{10.1103/PhysRevLett.106.048702}{ {PRL}}\
  }%
  \textbf{\bibinfo {volume} {106}},\ \bibinfo {pages} {048702} (
  \bibinfo {year} {2011});\ 
  \bibAnnoteFile{NoStop}{Roudi:2011p13014}%
  \bibfield{author}{%
  \bibinfo {author} {\bibfnamefont{J.~C.}\ \bibnamefont{Vasquez}}, \bibinfo
  {author} {\bibfnamefont{O.}~\bibnamefont{Marre}}, \bibinfo {author}
  {\bibfnamefont{A.~G.}\ \bibnamefont{Palacios}}, \bibinfo {author}
  {\bibfnamefont{M.~J.~Berry II}},\ and\ \bibinfo {author}
  {\bibfnamefont{B.}~\bibnamefont{Cessac}},\ }%
  \bibfield{journal}{%
  \Doi{10.1016/j.jphysparis.2011.11.001}{\bibinfo {journal} {J.
  Physiol. Paris}}\ }%
  \textbf{\bibinfo {volume} {106}},\ \bibinfo {pages} {120} (
\bibinfo {year} {2012})
  \bibAnnoteFile{NoStop}{Vasquez:2012p13088}%
\bibitem{Parrish}%
  \BibitemOpen
  \bibfield{author}{%
  \bibinfo {author} {\bibfnamefont{J.~K.}\ \bibnamefont{Parrish}}\ and\
  \bibinfo {author} {\bibfnamefont{W.~M.}\ \bibnamefont{Hamner}},\ }%
  \emph{\bibinfo {title} {Animal Groups in Three Dimensions}}\ (\bibinfo
  {publisher} {Cambridge University Press},\ \bibinfo {address} {Cambridge},\
  \bibinfo {year} {1997})%
  \bibAnnoteFile{NoStop}{Parrish}%
\bibitem{twoblock}%
  \BibitemOpen
  \bibfield{author}{%
  \bibinfo {author} {\bibfnamefont{H.~P.}\ \bibnamefont{Zhang}}, \bibinfo
  {author} {\bibfnamefont{A.}~\bibnamefont{Be'er}}, \bibinfo {author}
  {\bibfnamefont{E.-L.}\ \bibnamefont{Florin}},\ and\ \bibinfo {author}
  {\bibfnamefont{H.~L.}\ \bibnamefont{Swinney}},\ }%
  \bibfield{journal}{%
  \Doi{10.1073/pnas.1001651107}{\bibinfo {journal} {PNAS}}\
  }%
  \textbf{\bibinfo {volume} {107}},\ \bibinfo {pages} {13626} (
  \bibinfo {year} {2010});\ %
  \bibAnnoteFile{NoStop}{Zhang:2010p12841}%
  \bibfield{author}{%
  \bibinfo {author} {\bibfnamefont{X.}~\bibnamefont{Chen}}, \bibinfo {author}
  {\bibfnamefont{X.}~\bibnamefont{Dong}}, \bibinfo {author}
  {\bibfnamefont{A.}~\bibnamefont{Be'er}}, \bibinfo {author}
  {\bibfnamefont{H.~L.}\ \bibnamefont{Swinney}},\ and\ \bibinfo {author}
  {\bibfnamefont{H.~P.}\ \bibnamefont{Zhang}},\ }%
  \bibfield{journal}{%
  \bibinfo {journal} {PRL}\ }%
  \textbf{\bibinfo {volume} {108}},\ \bibinfo {pages} {148101} (
  \bibinfo {year} {2012})%
  \bibAnnoteFile{NoStop}{Chen:2012p12842}%
\bibitem{Sumino}%
  \BibitemOpen
  \bibfield{author}{%
  \bibinfo {author} {\bibfnamefont{Y.~Sumino.}\ \bibnamefont{{\it et al.}}},\ }%
  \bibfield{journal}{%
  \bibinfo {journal} {Nature (London)}\ }%
  \textbf{\bibinfo {volume} {483}},\ \bibinfo {pages} {448} (\bibinfo {year}
  {2012})%
  \bibAnnoteFile{NoStop}{Sumino}%
\bibitem{Sepulveda:2013p13003}%
  \BibitemOpen
  \bibfield{author}{%
  \bibinfo {author} {\bibfnamefont{N.}~\bibnamefont{Sep{\'u}lveda}}, \bibinfo
  {author} {\bibfnamefont{L.}~\bibnamefont{Petitjean}}, \bibinfo {author}
  {\bibfnamefont{O.}~\bibnamefont{Cochet}}, \bibinfo {author}
  {\bibfnamefont{E.}~\bibnamefont{Grasland-Mongrain}}, \bibinfo {author}
  {\bibfnamefont{P.}~\bibnamefont{Silberzan}},\ and\ \bibinfo {author}
  {\bibfnamefont{V.}~\bibnamefont{Hakim}},\ }%
  \bibfield{journal}{%
  \Doi{10.1371/journal.pcbi.1002944.s014}{\bibinfo {journal} {PLoS Comput
  Biol}}\ }%
  \textbf{\bibinfo {volume} {9}},\ \bibinfo {pages} {e1002944} (
 \bibinfo {year} {2013})%
  \bibAnnoteFile{NoStop}{Sepulveda:2013p13003}%
\bibitem{twoblock2}%
  \BibitemOpen
  \bibfield{author}{%
  \bibinfo {author} {\bibfnamefont{J.}~\bibnamefont{Deseigne}}, \bibinfo
  {author} {\bibfnamefont{O.}~\bibnamefont{Dauchot}},\ and\ \bibinfo {author}
  {\bibfnamefont{H.}~\bibnamefont{Chat{\'e}}},\ }%
  \bibfield{journal}{%
  \bibinfo {journal} {PRL}\ }%
  \textbf{\bibinfo {volume} {105}},\ \bibinfo {pages} {098001} (
  \bibinfo {year} {2010});\ %
  \bibAnnoteFile{NoStop}{Deseigne:2010p12917}%
  \bibfield{author}{%
  \bibinfo {author} {\bibfnamefont{C.}~\bibnamefont{Weber}}, \bibinfo {author}
  {\bibfnamefont{T.}~\bibnamefont{Hanke}}, \bibinfo {author}
  {\bibfnamefont{J.}~\bibnamefont{Deseigne}}, \bibinfo {author}
  {\bibfnamefont{S.}~\bibnamefont{L{\'e}onard}}, \bibinfo {author}
  {\bibfnamefont{O.}~\bibnamefont{Dauchot}}, \bibinfo {author}
  {\bibfnamefont{E.}~\bibnamefont{Frey}},\ and\ \bibinfo {author}
  {\bibfnamefont{H.}~\bibnamefont{Chat{\'e}}},\ }%
  \bibfield{journal}{%
  {PRL}\ }%
  \textbf{\bibinfo {volume} {110}},\ \bibinfo {pages} {208001} (\bibinfo {year}
  {2013})%
  \bibAnnoteFile{NoStop}{Weber:2013p13046}%
\bibitem{rods}%
  \BibitemOpen
  \bibfield{author}{%
  \bibinfo {author} {\bibfnamefont{F.}~\bibnamefont{Ginelli}}, \bibinfo
  {author} {\bibfnamefont{F.}~\bibnamefont{Peruani}}, \bibinfo {author}
  {\bibfnamefont{M.}~\bibnamefont{B\"{a}r}},\ and\ \bibinfo {author}
  {\bibfnamefont{H.}~\bibnamefont{Chat\'{e}}},\ }%
  \bibfield{journal}{%
  {PRL}\ }%
  \textbf{\bibinfo {volume} {104}},\ \bibinfo {pages} {18452} (\bibinfo {year}
  {2010})%
  \bibAnnoteFile{NoStop}{rods}%
\bibitem{Attanasi:2013p12927}%
  \BibitemOpen
  \bibfield{author}{%
  \bibinfo {author} {\bibfnamefont{A.}~\bibnamefont{Attanasi~et~al.}},} 
  \bibfield{journal}{%
  \bibinfo {journal} {arXiv:1303.7097v1}\ }%
  (\bibinfo {year} {2013})
  \bibAnnoteFile{NoStop}{Attanasi:2013p12927}%
\end{thebibliography}%

\section{Appendix}

\subsection*{Maximum entropy approach}

In the maximum entropy approach, one looks for the maximally
disordered probability distribution consistent with carefully
chosen observables of the data. In practice, given a stochastic variable
$\mathbf{s}$, and a set of observables
$\{\mathcal{O}_{\mu}(\mathbf{s})\}$, with $\mu=1,\ldots,K$, one
looks for the model distribution $P$ of maximum entropy
\beq
S[P]=-\sum_{\mathbf{s}}P(\mathbf{s})\ln P(\mathbf{s}),
\eeq
that coincides with the data for the average values of each of the observables:
\beq\label{eq:constraints}
\<\mathcal{O}_{\mu}\>_{\rm data}=\<\mathcal{O}_{\mu}\>_{P}.
\eeq
Using the technique of Lagrange multipliers, one shows that the
distribution takes the exponential form:
\beq\label{eq:maxent}
P(\mathbf{s})=\frac{1}{\mathcal{Z}(\{\lambda_\mu\})}\exp\left(-\sum_{\mu=1}^{K}\lambda_\mu
  \mathcal{O}_\mu(\mathbf{s})\right),
\eeq
where $\{\lambda_\mu\}$ are Lagrange multipliers that need to be
set to satisfy \eqref{eq:constraints}, and
$\mathcal{Z}(\{\lambda_\mu\})$ is a normalization factor enforcing
$\sum_{\mathbf{s}} P(\mathbf{s})=1$. By analogy with the Boltzman
distribution from equilibrium
statistical mechanics, the sum inside the exponential may be
interpreted as an energy.

Conveniently, the Lagrange multipliers that match the mean value of the observables
are also those that maximize the likelihood of the data given the
exponential form
\eqref{eq:maxent}. Given $M$ data points
$\mathbf{s}^1,\ldots,\mathbf{s}^M$, the log-likelihood of the data
reads:
\beq
\begin{split}
&\ln\mathcal{P}(\{\lambda_\mu\})\equiv\ln\prod_{a=1}^M P(\mathbf{s}^a)\\
&\quad=-\sum_{a=1}^M \sum_{\mu=1}^K
\lambda_\mu\mathcal{O}_\mu(\mathbf{s}^a) -M\ln
\mathcal{Z}(\{\lambda_\mu\}).
\end{split}
\eeq
Maximizing the log-likelihood with respect to the parameters
$\{\lambda_\mu\}$ implies:
\beq
\begin{split}
\frac{\partial\ln\mathcal{P}(\{\lambda_\mu\})}{\partial \lambda_\mu}=M\lb -\frac{\partial
  \ln\mathcal{Z}}{\partial \lambda_\mu} -
\<\mathcal{O}(\mathbf{s})\>_{\rm data}\rb&=0\\
M\lb \<\mathcal{O}(\mathbf{s})\>_{P}-\<\mathcal{O}(\mathbf{s})\>_{\rm
  data}\rb &=0.
\end{split}
\eeq
By virtue of this equivalence, we will maximize the expression of
the log-likelihood with respect to the parameters to find the correct
maximum entropy distribution.

Let us now consider the specific case of bird flocks.
Denote $\mathbf{s}=(s_1,\ldots,s_N)$ the flight directions of birds in
a flock of size $N$. The maximum entropy distribution consistent with the
{\em synchronous} pairwise correlation functions $\<s_is_j\>$, for all
$(i,j)$, reads:
\beq\label{eq:staticSI0}
P( \mathbf{s} )=\frac{1}{Z}\exp\lp \frac{1}{2}\sum_{ij}J^{\rm stat}_{ij}s_i s_j \rp,
\eeq
where $\{J_{ij}\}$ are (minus) the Lagrange multipliers associated to
the constraints on the correlation functions.

Generalizing the set of constrained obsersables to both synchronous and consecutive-time correlation
functions, $\{s_i^{t}s_j^t\}$ and $\{s_i^{t+1}s_j^t\}$, for all pair
$(i,j)$, and for all times $t$ in the trajectory, yields a
time-dependent maximum entropy distribution:
\beq\label{eq:maxentdynSI}
P( \mathbf{s}^1,\ldots, \mathbf{s}^T )=\frac{1}{\hat Z}\exp\lp - \mathcal{A}
\rp.
\eeq
with $\hat Z$ again a normalization factor, and
\beq\label{eq:action}
\mathcal{A}=-\frac{1}{2}\sum_t \sum_{i\neq j} \lp J^{(1)}_{ij;t} s_i^t s_j^t +
J^{(2)}_{ij;t} s_i^{t+1} s_j^t\rp,
\eeq
where $\{J^{(1)}_{ij;t}\}, \{J^{(2)}_{ij;t}\}$ are the Lagrange
multipliers associated to the constraints on the synchronous and
consecutive-time correlation functions.
Here, $\mathcal{A}$ is more appropriately interpreted as
an action, in a path-integral representation of the stochastic
trajectories of the whole flock.

\subsection{Markovian description}

Because the action only involves cross-terms between consecutive
times, it underlies a Markov process
\beq
P(\mathbf{s}^1,\ldots,\mathbf{s}^T)=P(\mathbf{s}^1)\prod_{t=1}^{T-1}P(\mathbf{s}^{t}|\mathbf{s}^{t-1})
\eeq
and can be rewritten as:
\beq\label{eq:AMarkov}
\mathcal{A}+\ln \hat Z=-\ln P(\mathbf{s}^1)+\sum_t \mathcal{L}_t(\mathbf{s}^{t+1},\mathbf{s}^t),
\eeq
where
\beq
\mathcal{L}_t(\mathbf{s}^{t+1},\mathbf{s}^t)\equiv-\ln P(\mathbf{s}^{t+1}|\mathbf{s}^t)
\eeq
may be interpreted as a Lagrangian density in the path integral
formalism.

Let us check that this Markovian decomposition is possible.
Identifying the two expressions of $\mathcal{A}$ \eqref{eq:action} and
\eqref{eq:AMarkov}, we may write $\mathcal{L}_t$ in the form:
\beq\label{eq:L}
\mathcal{L}_t(\mathbf{s}',\mathbf{s}) = -\frac{1}{2}\sum_{ij}
\left(J^{(2)}_{ij;t} s'_i s_j +
J^{(1)}_{ij;t} s_i s_j\right)
- K_{t}(\mathbf{s}') + K_{t-1}(\mathbf{s}),
\eeq
with the constraint that, for all $\mathbf{s}$, the transition
probability be normalized,
\beq
1=\sum_{\mathbf{s}'}\exp\lb-\mathcal{L}_t(\mathbf{s}',\mathbf{s})\rb
\eeq
which entails:
\beq\label{eq:rec2}
K_{t-1}(\mathbf{s})=\ln\sum_{\mathbf{s}'} \exp\lb\frac{1}{2}\sum_{ij}\lp
J^{(2)}_{ij;t} s'_i s_j+
J^{(1)}_{ij;t} s_i s_j\rp
+K_{t}(\mathbf{s}')\rb.
\eeq
Eq.~\eqref{eq:rec2} defines a descending
recursion, by which $K_t$ is calculated from the
next time point. Thus the Markovian form of the action is fully
specified using \eqref{eq:L}.

\subsection{Equivalence with a generalized Vicsek model in the spin-wave approximation}

In general, the integral in \eqref{eq:rec2} cannot be calculated
analytically, and $K_t$ does not have a simple quadratic form
as a function of $\mathbf{s}$. However things simplify
in the spin-wave approximation, where
the flock is very polarized, as we will show now. Denote $s_i=\pi_i + n\sqrt{1-(\pi_i)^2}$, where $n$ is
an abitrary unit vector, and $\pi_i$ is the perpendicular component of the orientation,
$\pi_i n=0$. $n$ is chosen to be close the flock's main direction of flight, so that $\pi_i\ll 1$.
Let us assume a quadratic form for $K_t$:
\beq
K_{t}(\mathbf{s})=\frac{1}{2}\sum_{ij} K_{ij;t} s_i s_j+U_t.
\eeq
The integral in \eqref{eq:rec2} can be expanded at small $\pi$:
\beq
\begin{split}
&K_{t-1}({\bm \pi})=\frac{1}{2}\sum_{ij}J^{(1)}_{ij;t} (1+\pi_i \pi_j-\pi_i^2)\\
&\quad+U_{t}+\frac{1}{2}\sum_{ij}\lp J^{(2)}_{ij;t}+K_{ij;t}\rp-\frac{1}{4}\sum_{ij}J^{(2)}_{ij;t}\pi_j^2,
\\
&\quad+\ln\int d{\bm \pi}' \exp\lb -\frac{1}{2}\sum_{ij}A_{ij;t} \pi'_i \pi'_j +\frac{1}{2}\sum_{ij}J^{(2)}_{ij;t}\pi'_i\pi_j\rb
\end{split}
\eeq
with
\beq\label{eq:defA}
A_{ij;t}=-K_{ij;t}+\delta_{ij}\sum_kK_{ik;t} +\frac{1}{2}\delta_{ij}\sum_kJ^{(2)}_{ik;t}.
\eeq
This Gaussian integral 
can be calculated exactly. Doing so, and
expanding the left-hand side of \eqref{eq:rec2} at small $\pi$, yields
\beq\label{eq:rec3}
\begin{split}
&{K}_{ij;t-1}-\delta_{ij}\sum_k
K_{ik;t-1}=J^{(1)}_{ij;t}-\delta_{ij}\sum_k J^{(1)}_{ik;t}\\
&\quad+\frac{1}{4}\lb \mathbf{J}^{(2)\dagger}_t
\mathbf{A}_t^{-1} \mathbf{J}^{(2)}_t\rb_{ij}-\frac{1}{2} \delta_{ij}
\sum_{k}J^{(2)}_{ik;t},
\end{split}
\eeq
\beq
\begin{split}
&U_{t-1}+\frac{1}{2}\sum_{ij}K_{ij;t-1}=\frac{1}{2}\sum_{ij}J^{(1)}_{ij;t}+U_{t} \\
&\qquad+\frac{1}{2}\sum_{ij}\lp
J^{(2)}_{ij;t}+H_{ij;t}\rp-\frac{d-1}{2}\ln{\lp \frac{\det
    \mathbf{A}_t}{(2\pi)^N}\rp}.
\end{split}
\eeq
Focusing on the non-diagonal terms of the matrix $\mathbf{K}_t$, we obtain
a simple expression for the recursion:
\beq
\mathbf{K}_{t-1}= \mathbf{J}^{(1)}_t+\frac{1}{4}{ \mathbf{J}^{(2)\dagger}_t} \mathbf{A}_t^{-1} \mathbf{J}^{(2)}_t.
\eeq

We can now replace the expression of $K_t$ 
\eqref{eq:L}, and thus rewrite the transition probability in terms of
$\pi$ in a Gaussian form:
\beq\label{eq:Lt}
\begin{split}
\mathcal{L}_t({\bm \pi'},{\bm \pi})=&-\frac{d-1}{2}\ln{\lp \frac{\det
    \mathbf{A}_t}{(2\pi)^N}\rp}\\
&{ +\frac{1}{2} \lp {\bm\pi}' -
\mathbf{M}_t{\bm\pi}\rp^{\dagger}\mathbf{A}_t \lp {\bm\pi}' -
\mathbf{M}_t{\bm\pi}\rp},
\end{split}
\eeq
with
\beq
\mathbf{M}_t=\frac{1}{2}\mathbf{A}_t^{-1} \mathbf{J}^{(2)}_t.
\eeq
This transition probability rule describes a random walk in the joint space
of bird directions, described by:
\beq\label{eq:rw}
\pi_i^{t+1}=\sum_j M_{ij;t} \pi_j^t + \epsilon_i^t,
\eeq
with $ {\bm \epsilon}^t$ a random, isotropic Gaussian noise perpendicular to $n$,
of zero mean and covariance: 
\beq\label{eq:noisecov}
\<{\bm\epsilon}^{t} ( {\bm
  \epsilon}^{t'})^{\dagger}\>=(d-1)\mathbf{A}_t^{-1}\delta_{t,t'}.
\eeq
Note that the $(d-1)$ factor, here and in previous equations,
corresponds to the dimensionality of the perpendicular component
$\pi$.

$\mathbf{M}_t$ defines a well-balanced weighted average, as it
satisfies:
\beq\label{eq:balanced}
\sum_{j}M_{ij;t}=1.
\eeq

To show this, let us rewrite this identity in a matrix form:
\beq
\frac{1}{2}\mathbf{A}_t^{-1}\mathbf{J}_t^{(2)} \mathbf{u}=\mathbf{u}
\eeq
where $\mathbf{u}$ is a vector of ones, $u_i=1$. Proving
\eqref{eq:balanced} is therefore equivalent to showing:
$
\mathbf{J}_t^{(2)}\mathbf{u} = 2 \mathbf{A}_t\mathbf{u}
$,
which follows from the definition of $\mathbf{A}_t$
\eqref{eq:defA}.

This identity also allows us to check that the diagonal components in the equality
\eqref{eq:rec3} are consistent with the off-diagonal
components. This is done by checking that on both sides of the \eqref{eq:rec3},
contraction with $\mathbf{u}$ gives zero.

The equation describing the collective random walk in
terms of the perpendicular component $\pi$ holds almost the same for the flight
direction $s$ itself. Starting from the update equation:
\beq
s^{t+1}_i=\theta\lb \sum_j M_{ij;t} s^t_j + \eta_i^t \rb,
\eeq
where $\theta(x) = x/\Vert x \Vert$ is the normalization operator, and
expanding in the spin-wave approximation ($\pi_i\ll 1$), one recovers \eqref{eq:rw}
with $\epsilon_i^t=\eta_i^t-(n\cdot\eta_i^t)n$ the
perpendicular component of the vectorial noise $\eta$.

\subsection*{Parametrization}

The matrices $\mathbf{M}_t$ and $\mathbf{A}_t$ are parametrized as
follows:
\beq\label{eq:par1}
M_{ij}=(1-J\delta t n_i)\delta_{ij}
+ J\delta t n_{ij},
\eeq
where $n_{ij}=1$ if $j$ is one of $i$'s neighbours, and 0
otherwise, and $n_i=\sum_{ij}n_{ij}$. (We drop the $t$ index, even
though $n_{ij}$ depends on $t$ in general); and
\beq\label{eq:par2}
A_{ij}=[1/(2\delta
tT)]\delta_{ij}.
\eeq
$J$ is interpreted as an alignment strength, and $T$ as a temperature.

\subsection*{Continuous time limit and equivalence with static maximum
entropy}

The parametrization has a well defined continuous-time limit. When
$\delta t\to 0$, \eqref{eq:rw}:
\beq\label{eq:langevin}
\frac{d {\bm\pi}}{dt}=-J{\bm \Lambda}{\bm \pi}+{\bm \xi}(t),
\eeq
where $\Lambda_{ij}=n_i\delta_{ij}-n_{ij}$, and $\xi_i(t)$ are i.i.d
Gaussian white noises with $\<\xi_i(t)\xi_i(t')\>=2T(d-1)
\delta(t-t')$, where $\delta(x)$ is Dirac's delta function.

When ${\bm \Lambda}$ varies slowly with time, \eqref{eq:langevin} can be
formally integrated:
\beq
{\bm \pi}(t)=\int_{-\infty}^t  dt'\, e^{-J{\bm
    \Lambda}(t-t')}{\bm \xi}(t')
\eeq
If, in addition, ${\bm \Lambda}$ is symmetric, the system reaches some
equilibrium steady state. More precisely, the collective mode that is parallel to
$\mathbf{u}$, which corresponds to the average direction of the flock
$(1/N)\sum_i \pi_i$, follows an unconstrained random walk, as it
corresponds to a the
zero mode of ${\bm \Lambda}$, ${\bm \Lambda}\mathbf{u}=0$. All the
other modes that are orthogonal to $\mathbf{u}$ are bounded by a
restoring force.
The steady-state
distribution of ${\bm \pi}$ is therefore Gaussian,
with $C_{ij}={\rm Cov}(\pi_i,\pi_j)$ satisfying:
\beq\label{eq:cov}
J{\bm \Lambda}\mathbf{C}=(d-1) T
\lp\mathbf{1}-\frac{\mathbf{u}\mathbf{u}^\dagger}{N}\rp,
\eeq
where $\mathbf{1}$ is the identity matrix.

Remarkably, in the spin-wave approximation, this distribution is the
same as the one obtained by the principle
maximum entropy constrained by the static correlation functions:
\beq\label{eq:staticSI}
P( \mathbf{s} )=\frac{1}{Z}\exp\lp \frac{1}{2}\sum_{i\neq j}J^{\rm stat}_{ij}s_i s_j \rp,
\eeq
with
\beq
J^{\rm stat}_{ij}=\frac{J}{T} n_{ij}.
\eeq
One can check this by expanding \eqref{eq:staticSI} at small $\pi$, after setting $n$ to be
the average direction of the flock, so that $\sum \pi_i=0$, and
\beq
P({\bm \pi})\propto \delta\lp \sum_i \pi_i\rp \exp\lp -\frac{J}{2T} \sum_{ij} \Lambda_{ij} \pi_i
\pi_j\rp.
\eeq
By virtue of Gaussian integration rules, this distribution has the same 
covariance as \eqref{eq:cov}, and therefore is identical.

\subsection*{Parameter inference}

We can rewrite the Lagrangian \eqref{eq:Lt}  in a slightly different manner:
\beq\label{eq:Lt2}
\begin{split}
\mathcal{L}_t( {\bm \pi}^{t+1}| {\bm \pi}^t)
&=-\frac{d-1}{2}\ln{\lp \frac{\det
    \mathbf{A}_t}{(2\pi)^N}\rp} +\frac{1}{2}\mathrm{Tr}( \mathbf{C}_{t+1} \mathbf{A}_{t}^{\dagger})\\
&-\frac{1}{2}\mathrm{Tr}( \mathbf{J}^{(2)}_{t} \mathbf{G}_{t}^{\dagger})
+\frac{1}{8}\mathrm{Tr}({ \mathbf{J}^{(2)}_t}^{\dagger} \mathbf{A}_t^{-1} \mathbf{J}^{(2)}_t \mathbf{C}_t ^{\dagger}),
\end{split}
\eeq
where $ \mathbf{C}_t= {\bm \pi}^t ( {\bm \pi}^{t})^{\dagger}$ and $ \mathbf{G}_t= {\bm \pi}^{t+1}
( {\bm \pi}^{t})^{\dagger}$.

Under the parametrization \eqref{eq:par1},\eqref{eq:par2}, the
minus-log-likelihood \eqref{eq:Lt2} becomes (the time index is implicit from now on):
\beq\label{eq:Lvicsek}
\frac{\mathcal{L}}{N}=\frac{d-1}{2}\ln 2T\delta t +\frac{\mathcal{\hat
    L}}{4T\delta t},
\eeq
where
\beq
\begin{split}
\mathcal{\hat L}=&C_s^1+C_s-2\alpha \tilde C_s+\alpha^2\hat
C_s+2\alpha(C_{\rm int}-\alpha \tilde C_{\rm int})\\
&+\alpha^2 C'_{\rm int}-2\alpha G_{\rm int}
-2(G_s-\alpha \tilde G_s),
\end{split}
\eeq
and $\alpha=Jn_c \delta t$.

The various correlated functions used in this expression are defined
in Table I in the main text.

In the case of non-constant $n_i$, $n_c$ is defined as
$(1/N)\sum_i n_i$.
Note that in the case of constant $n_i=n_c$, as in the case of the
nearest-neighbor model, $\tilde C_s=\hat C_s=C_s$, $\tilde G_s=G_s$
and $\tilde C_{\rm int}=C_{\rm int}$.

There are three parameters to optimize over: the interaction strengh
$J$, the interaction range $n_c$, and the ``temperature'' $T$ which
sets the strength of noise. This last one is simply given by the
condition $\partial \mathcal{L}/\partial T=0$, which yields:
\beq\label{eq:Tstar}
T=\frac{\mathcal{\hat L}}{2(d-1)\delta t}.
\eeq
At this optimum value of $T$, we have 
\beq
\frac{\mathcal{L}}{N}=\frac{d-1}{2} \{\ln[\mathcal{\hat L}/(d-1)]+1\}.
\eeq
Minimizing $\mathcal{\hat L}$, $\partial \mathcal{\hat L}/\partial \alpha$, then yields the optimum value of $\alpha$:
\beq\label{eq:alphaSI}
\alpha=\frac{C_{\rm int}-\tilde C_s+\tilde G_s -G_{\rm int}}{2\tilde C_{\rm int}-C'_{\rm
    int} -\hat C_s}.
\eeq
At this optimum, one has $\mathcal{\hat
  L}=C_s^1+C_s-2G_s+\mathcal{\tilde L}$,
where
\beq
\mathcal{\tilde L}=\frac{\lp C_{\rm int}-\tilde C_s+\tilde G_s -G_{\rm int}\rp^2}{2\tilde C_{\rm int}-C'_{\rm
    int} -\hat C_s}
\eeq
is the only term that depends on the interaction matrix $n_{ij}$.
Therefore, to find the optimum interaction range $n_c$ in the case of the
nearest-neighbor model, one just needs to minimize
${\mathcal{\tilde L}}(n_c)$.

\subsection*{Consistency with the static approach}
To recover the static inference equations, we start by rewriting the dynamical inference equations, Eqs.~\eqref{eq:Tstar} and~\eqref{eq:alphaSI}, explicitly:
\bea
    J&=&\frac{1}{n_c}\frac{\Omega + (d-1)T_0}
    {C'_{\rm int}+\hat{C}_s-2\tilde{C}_{\rm
        int}},\label{eq:inferenceinSI}\\
T &=& T_0+\frac{C_s^1-C_s}{2(d-1)\delta t} - \frac{J\,n_c\delta
  t}{2(d-1)}\left(\frac{\tilde C_s-\tilde G_s}{\delta t}+\Omega\right),\label{eq:TinSI}
\eea
with $n_c=(1/N)\sum_i n_i$ and 
\beq\label{eq:T0OmegaSI}
T_0=\frac{{ C}_s-{ G}_s}{\delta t(d-1)}, \quad \Omega=\frac{G_{\rm int}-C_{\rm int}}{\delta t}.
\eeq
When the system is at steady state, we have $C_s^1 \!\! \approx \!\!
C_s$ and
$\Omega \approx (2Nn_c)^{-1}\sum_{ij}
\pi_i\pi_j \frac{dn_{ij}}{dt}$ (directly from definitions in Table I
of the main text and Eq.~\ref{eq:T0OmegaSI}); 
the
second term in Eq.~\eqref{eq:TinSI} cancels and $T  \approx T_0$ for small
$\delta t$. If we further assume that data was actually generated by
{\em exactly} the class of models we are trying to infer (which may
not be the case in general, as we are looking at effective
descriptions), we have exactly $T = T_0$. If in addition neighbor changes are slow, then
$\Omega  \approx  0$ and Eq.~(\ref{eq:cov}) implies $\tilde C_{\rm int}
\! \approx \! C'_{\rm int}$. Eq.~(\ref{eq:inferenceinSI}) thus gives
\beq
\frac{Jn_c}{T} \approx \frac{d-1}{\hat C_s-\tilde C_{\rm int}},
\label{static}
\eeq
which is the result of the static inference \cite{Bialek:2012p12539}. Note however that in addition to recovering the alignment strength,
the dynamical inference procedure allows us to separate the
interaction coupling $J$ from the temperature $T$.

\subsection*{Spin wave expansion of the Topological Vicsek model}

As described in the main text, to test our dynamical inference method
we generated synthetic data with the Topological VM defined by
\bea
\label{VM1SI}
\theta_i^{t+\delta t} &=&\mathrm{Arg}[ s_i^t + J_V\delta t \sum_j
n_{ij}s_j^t ]
+\sqrt{\delta t}\,\xi_i^t,\\
\label{VM2SI}
r_i^{t+\delta t}&=&r_i^t+v_0\,\delta t \, s_i^{t+\delta t}.
\eea
In this section, we show that Eq.~(\ref{VM1SI}) is in
fact equivalent in the spin-wave limit to an update equation of the
same kind as Eqs.~(\ref{eq:rw}),(\ref{eq:par1}) and (\ref{eq:par2}). To this aim, it is convenient to rewrite Eq.~(\ref{VM1SI}) in the following equivalent form 

\bea
\label{VM1app}
s_i^{t+\delta t} &=&  \frac{s_i^t + J_V\delta t \sum_j n_{ij} s_j^t}{\|s_i^t + J_V\delta t \sum_j n_{ij} s_j^t\|}
+\sqrt{\delta t}\,\epsilon_i^t,\\
\label{VM2app}
r_i^{t+\delta t}&=&r_i^t+v_0\,\delta t \, s_i^{t+\delta t},
\eea
where  $\epsilon_i$ is  a delta-correlated noise  perpendicular to $s_i$ with variance $2 (d-1) T_V$ (i.e. whose effect is the same as the angular noise appearing in Eq.~(\ref{VM1SI})).

In the large polarization regime we can perform a spin wave expansion
$s_i=\pi_i + n\sqrt{1-\pi_i^2}$, where $n$ is a vector representing  the  global direction of motion and  $\pi_i $ is the component of the direction $s_i$ perpendicular to $n$. We can now expand the normalization at the r.h.s. in Eq.~(\ref{VM1app}) with respect to $\pi_i^2$ to get
\beq
\|s_i^t + J_V\delta t \sum_j n_{ij} s_j^t\|  = 1+\delta t J_V n_i + \text{O}(\pi^2)
\label{expan}
\eeq
where $n_i=\sum_j n_{ij}$.  Eq.~(\ref{VM1app}) then leads to the following update equation for the $\{\pi_i\}$
\bea
\pi_i^{t+\delta t}&=&\frac{\pi_i^t+\delta t J_V \sum_j n_{ij}\pi^t_j} {1+\delta t J_V n_i}+\sqrt{\delta t} \epsilon_i
\phantom{cuccagnacuccagna}\nonumber\\
&=& \left (1-\delta t \frac{J_V}{1+\delta t J_V n_i}\right ) \pi_i^t \nonumber \\
&&\quad\quad + \delta t \frac{J_V}{1+\delta t J_V n_i}\sum_j n_{ij}\pi_j^t + \sqrt{\delta t}\epsilon_i
\label{VMexp}.
\eea
When $\delta t$ is small, we can disregard fluctuations in $n_i$ and Eq.~(\ref{VMexp}) is of the same form of Eqs.~(\ref{eq:rw}) with the parametrization defined in (\ref{eq:par1})-(\ref{eq:par2}) and
\beq
J=\frac{J_V}{1+\delta t J_V n_V}.
\eeq

\end{document}